# Large spin Hall magnetoresistance and its correlation to the spin-orbit torque in W/CoFeB/MgO structures

Soonha Cho[1], Seung-heon Chris Baek[1,2], Younghun Jo[3], and Byong-Guk Park[1,*]

[1] Department of Materials Science and Engineering, KAIST, Daejeon 305-701, Korea

[2] Department of Electrical Engineering, KAIST, Daejeon 305-701, Korea

[3] Division of Scientific Instrumentation KBSI, Daejeon 305-806, Korea

* corresponding email: bgpark@kaist.ac.kr (B.-G. Park)

**Abstract**

The spin-orbit interaction in heavy metal/ferromagnet/oxide structures has been extensively investigated because it can be employed in manipulation of the magnetization direction by in-plane current. This implies the existence of an inverse effect, in which the conductivity in such structures should depend on the magnetization orientation. In this work, we report a systematic study of the magnetoresistance (MR) of the W/CoFeB/MgO structures and its correlation to the current-induced torque to the magnetization. We observe that the MR is independent of the angle between magnetization and current direction, but is determined by the relative magnetization orientation with respect to the spin direction accumulated by spin Hall effect, which is the same symmetry of so-called spin Hall magnetoresistance. The MR of ~1% in W/CoFeB/MgO samples is considerably larger than those in other structures of Ta/CoFeB/MgO or Pt/Co/AlOx, which indicates a larger spin Hall angle of W. Moreover, the similar W thickness dependence of the MR and the current-induced magnetization switching efficiency demonstrates that they share the same underlying physics, which allows one to utilize the MR in non-magnet/ferromagnet structure in order to understand closely related other spin-orbit coupling effects such as inverse spin Hall effect, spin-orbit spin transfer torques, etc.

The spin Hall effect (SHE)[1,2], the generation of a spin current from a charge current in non-magnetic (NM) materials, has drawn an increasing interest because it can be utilized in the spintronic devices for current-induced magnetization switching[3-5] as well as for high speed domain wall motion[6,7]. In the ferromagnet (FM)/NM heterostructures, the SHE-induced spin accumulation interacts with the local magnetic moment in FM depending on their relative directions. When the spin orientation ($\vec{\sigma}$) is non-collinear to the magnetization direction ($\vec{M}$), the accumulated spins are absorbed by FM, which in return exerts a torque on the magnetic moment, whereas the spins are reflected when the $\vec{\sigma}$ is parallel to $\vec{M}$. The reflection of the spin current is added to the original charge current via inverse spin Hall effect (ISHE) as depicted in Fig. 1(a). Consequently, the resistance of the FM/NM structures depends on the relative orientations of the magnetization and accumulated spins, which is so-called spin Hall magnetoresistance (SMR) since it is based on the SHE as well as the ISHE in NM[8-10]. Thus, the magnitude of SMR is proportional to the square of the spin Hall angle. So far, SMR has been investigated in FM/NM systems with a magnetic insulator, mostly yttrium iron garnets (YIGs), in which other magnetoresistance (MR) effect is absent as the current only flows through NM[8,10-13]. However, SMR should be present even in fully metallic FM/NM structures because the SHE and ISHE are generic features of NM. Moreover, SMR is an inverse effect of the spin-orbit torques[13], in which the magnetization direction can be modulated by an in-plane current through the SHE and/or interfacial Rashba effect. Since the spin-orbit torques have been investigated mostly in fully metallic NM/FM structures[14-20], the study on SMR in metallic structures is of great importance to understand the spin transport as well as the origin of the spin-orbit torques in those structures.

In this study, we investigate SMR in W/CoFeB/MgO structures as a function of the thicknesses of W and CoFeB layers. We observe that a SMR is not sensitive to the CoFeB thickness, but is strongly dependent on the W thickness. A large SMR of about 1% is observed in W/CoFeB/MgO samples, which is about two orders of magnitude greater than that in the YIG/Pt samples[8,10-13], and also greater than those in Ta/CoFeB/MgO or Pt/Co/AlOx structures. We attribute this significant SMR in W-based structures to the large effective spin Hall angle of W[21,22]. The spin Hall angle and the spin diffusion length of W are 0.22±0.03 and 2.0±0.5 nm, respectively, which are extracted from the thickness dependence of SMR values. Moreover, we perform spin-orbit torque-induced magnetization switching experiments and find that the switching efficiency is closely correlated to the magnitude of SMR. This confirms that SMR and spin-orbit torques share the same underlying physics.

**Results**

**Spin Hall magnetoresistance.**

We first present measurement of the longitudinal ($R_{xx}$) and transverse resistances ($R_{xy}$) of W(5 nm)/CoFeB(1.2 nm)/MgO(1.6 nm) sample as a function of in-plane magnetic fields of $H_x$ and $H_y$ in Fig. 1 (b) and 1(c), respectively. As the sample has perpendicular magnetic anisotropy, the in-plane magnetic fields rotate the magnetization from out-of-plane (z-direction) to in-plane (x- or y- direction). We apply in-plane magnetic field up to 1.5 T, which is larger than the anisotropy field (~1T) of the sample. As shown in Fig. 1(b), $R_{xx}$ is strongly dependent on the direction of magnetic field. We observe that $R_{xx}$ is almost constant for $H_x$, applied along the current direction. The application of $H_x$ rotates $\vec{M}$ in the x-z plane, so that the angle between $\vec{M}$ and current $I_x$ varies from 90º to zero. Therefore, the insensitivity of $R_{xx}$ to $H_x$ demonstrates a negligible conventional AMR effect in this sample. On the contrary, $R_{xx}$ is gradually reduced for the application of $H_y$, in which $\vec{M}$ is always perpendicular to $I_x$. In this field geometry, $R_{xx}$ is sensitive to the relative angle of $\vec{M}$ with respect to the y-direction. This magnetoresistance can be attributed to SMR[8], in which the SHE-induced spins pointing the y-direction of NM interact with the local magnetic moment of FM depending on the relative angle between them. For $\vec{M} \parallel y$, the accumulated spins are totally reflected at the NM/FM interfaces and are transferred to an additional charge current via ISHE, resulting in a lower resistance. We note that SMR in W/CoFeB/MgO sample is ~1.15%, which is about two orders of magnitude greater than those reported in Pt/YIG structures[8,10-13]. On the other hand, the transverse resistance $R_{xy}$ decreases as the in-plane field is increased, irrespective of the field direction. This indicates that $R_{xy}$ is dominated by the anomalous Hall effect (AHE) so that the SMR contribution to the $R_{xy}$ is negligibly small, which, however is only valid for the magnetic field of $H_x$ or $H_y$. The general feature of the contribution of SMR to $R_{xy}$ will be discussed later in Fig. 2.

In order to confirm the angular dependence of $R_{xx}$ and $R_{xy}$, we repeat the measurement with rotating the samples in three major planes, x-y, y-z, and x-z plane at strong magnetic fields. As shown in Fig. 2 (a), the representative angles of each plane are denoted as α, β, and γ, respectively. We note that the magnetic field of 1.5T (8T) for α (β and γ) rotation is larger than the anisotropy field of ~1T. Fig. 2(b) shows that $R_{xx}$ varies significantly with α and β, but remains almost constant with γ. The longitudinal resistivity ($\rho_{xx}$) in the FM/NM bilayer structure can be expressed[9] by

$$\rho_{xx} = \rho + \Delta\rho_o + \Delta\rho_1(1 - m_y^2) + \Delta\rho_2(m_x^2), \quad (1)$$

where $\rho$ is the intrinsic electric resistivity, $\Delta\rho_o$ is the reduced resistivity by spin-orbit interaction, and $\Delta\rho_1$ ($\Delta\rho_2$) is the change in resistivity owing to SMR (AMR). The x and y component of the magnetization ($m_x$ and $m_y$) is equivalent to $cos\gamma$ and $cos\beta$, respectively. From fitting the angular dependence curves (Fig. 2(b)) using Eq. (1), we obtain $\Delta\rho_1$ of ~3.4 μΩ cm ($\Delta R$~850Ω in the measurement graph) and negligible $\Delta\rho_2$. Therefore, in the W/CoFeB/MgO samples, SMR is much more dominant than AMR.

On the other hand, $R_{xy}$ depends on the rotating angles with all three directions (Fig. 2(c)). The dependence of $R_{xy}$ on β and γ is due to the AHE, in which $R_{xy}$ is gradually reduced as $\vec{M}$ rotates toward in-plane, whereas the variation of $R_{xy}$ with α is attributed to the planar Hall effect (PHE). We note that PHE is normally a transverse component of AMR (i.e., transverse AMR). However in this sample, AMR is negligible, so that PHE can be attributed to the transverse SMR. This can explain the large PHE value observed in similar structures[23], which is comparable to that of the AHE.

We compare the $R_{xx}$'s of W/CoFeB/MgO samples with those of samples with different underlayers. Figure 3 shows $R_{xx}$ of Ta/CoFeB/MgO sample (a) and Pt/Co/AlOx sample (b) as a function of two different magnetic fields of $H_x$ (black solid symbols, $\vec{M} \parallel \vec{I}$) and $H_y$ (red open symbols, $\vec{M} \perp \vec{I}$). Both samples show the similar behavior as the W/CoFeB/MgO samples, i.e., a stronger dependence of $R_{xx}$ on $H_y$ rather than $H_x$, demonstrating that SMR is the dominant magnetotransport in these NM/FM/oxide structures. However, the magnitude of SMR of the sample with Ta or Pt is significantly smaller than that of W/CoFeB/MgO structure, even though it is still much larger than those of Pt/YIG samples[8,10-13]. Since the mechanism of the SMR is understood by a combination of the SHE and ISHE, the larger SMR can be explained by a larger spin Hall angle of W than that of Ta or Pt, which is consistent with the reported values in literature[4, 21,22,24,25]. Since the SMR is proportional to the square of spin Hall angle[9], a relative magnitude of SMR indicates that the spin Hall angle of W is about three (two) times larger than the sample with Ta (Pt).

**Thickness dependence of the spin Hall magnetoresistance.**

In order to understand the SMR of W/CoFeB/MgO samples in more detail, we investigate the dependence of SMR on the thicknesses of W and CoFeB layers. We first examine the effect of CoFeB thickness on SMR. Figure 4(a) shows $R_{xx}$ as a function of a transverse field, $H_y$ for samples with different CoFeB thicknesses ranging from 0.8 to 1.4 nm, in which perpendicular magnetic anisotropy can persist. We find that normalized $R_{xx}$ by the resistance at $H_y$=1.5 T ($R_0$) does not vary much with the CoFeB thickness. The similar angular dependence of those four samples shown in Fig. 4(b) confirms that $R_{xx}$ does not significantly rely on the CoFeB thickness. To verify the role of W layer, we compare samples of thicker CoFeB (3.0 nm) with and without W layer. Both samples show the in-plane magnetic anisotropy. Figure 4(c) shows $R_{xx}$ as a function of $H_y$. We note that the magnetization of both samples is aligned in the $x$-direction with no magnetic field because of the shape anisotropy of Hall bar structure. The SMR is considerably larger for the sample with W underlayer. The results shown in Fig. 4 clearly demonstrate that the W layer has a key role in the observed SMR.

Next we examine the dependence of SMR on the W thickness in W(2~7 nm)/CoFeB(1.0 nm)/MgO(1.6 nm) samples. As shown in Fig. 5(a), the normalized $R_{xx}$ is the largest for W of 4 or 5 nm and is reduced for a thicker or thinner W layer. This strong thickness dependence supports the argument that MR in W/CoFeB/MgO structures is dominated by the SHE in W. The SMR can be expressed by an equation[9,10]

$$\frac{\Delta\rho}{\rho} = \theta_{SH}^2 \frac{\lambda}{d_N} Re\left(\frac{2\lambda G_{\uparrow\downarrow} tanh^2 \frac{d_N}{\lambda}}{\sigma + 2\lambda G_{\uparrow\downarrow} coth^2 \frac{d_N}{\lambda}}\right), \qquad (2)$$

where the $\theta_{SH}$, $\lambda$, $d_N$ is spin Hall angle, spin diffusion length, and the thickness of the NM layers, respectively. The $\sigma = \rho^{-1}$ is conductivity and $Re\ G_{\uparrow\downarrow}$ is real part of the spin mixing conductance. According to Eq. (2), SMR is decreased when the W layer is thinner than the spin diffusion length, because of the reduced spin current by back reflection at the interface. On the other hand, for a thicker W layer, the SMR is also reduced by a current shunting effect. From fitting of the thickness dependence of the SMR to the Eq. (2) (red line in Fig. 4(b)), the spin Hall angle of 0.22±0.03, and spin diffusion length of 2.0±0.5 can be extracted. Moreover the spin mixing conductance of the W/CoFeB is obtained to be 4.5±1.0 × $10^{14} \Omega^{-1} m^{-1}$. The data for the samples with larger thickness of W (> 6 nm) are largely deviated from the curve, which can be explained by the fact that the crystallographic structure of a thicker W layer might be different from a thinner one[21]. This is evidenced by a large drop of the resistivity for W layer of thickness larger than 6 nm (Fig. 5(c)). The resistance is inversely proportional to the thickness up to 5 nm, which gives the resistivity of ~300 μΩ cm. This high resistivity value indicates that the W layer is β-phase when it is thinner than 5 nm. However, it starts to deviate from the linear fit at larger thicknesses, implying the development of a different phase with a lower resistivity. This explains the large disagreement between the measured SMR and the theoretical prediction in Fig. 5(b) at thicker W cases.

**Correlation of the spin Hall magnetoresistance with spin-orbit torque.**

Hitherto, we study the transport characteristics of W/CoFeB/MgO samples. Now, we examine the inverse effect of SMR, i.e., in-plane current-induced spin-orbit torque (SOT). In order to evaluate the magnitude of SOT, we perform the switching experiment using the same samples shown in Fig. 5. We first initialize the magnetization in the +z direction ($M_z = +1$), and sweep a pulsed current with pulse width of 10 μs from a positive to a negative value, and vice versa, while keeping a longitudinal magnetic field $H_x$ of 200 Oe, which is necessary for deterministic switching[3,5]. After each current pulse, the magnetization direction is detected by measuring AHE voltage. When the applied negative pulsed current is larger than a certain threshold, the reversal of the magnetization from +z to −z direction is observed. Note that a negative current and a positive $H_x$ favors the -z direction of magnetization ($M_z = -1$), which corresponds to the SOT with a negative spin Hall angle. We repeat the switching experiments for samples with various W thicknesses as shown in Fig. 6(a). We find that the critical current density for magnetization switching strongly depends on the W thickness. For example, , the switching can be done at the current density of ~11 MA/cm$^2$ for the samples with 5 nm W, while it exceeds 42 MA/cm$^2$ for that with 7 nm W. In order to compare the magnitude of SOT from the switching experiment, the ratio of the critical current density ($I_c$) and the magnetic anisotropy ($H_k$) is plotted in Fig. 6 (b), as this ratio is a rough estimate of SOT strength[26]. This shows that the ratio ($I_c/H_k$) is the minimum at W of 5 nm, where SMR is the maximum (see Fig. 5(b)). Since the ratio corresponds to the inverse of SOT efficiency, it indicates that the SOT magnitude has the same W thickness dependence as that of SMR. It suggests that the SMR and the SOT share the same physical origin of the SHE.

**Discussions**

The W/CoFeB/MgO structures show considerably larger SMR than structures with FM insulator or FM metal with other NM materials. This can be attributed to a large spin Hall angle of W since the SMR originates from the SHE in NM. However, we note that there may be some contributions from FM or FM/NM interface as well. The SMR reported in most literature utilizing Pt/YIG structure[8,10-13] is about 0.01%, which is much smaller than the result from Pt/Co/AlOx structure shown in Fig. 3(b). Assuming the same spin Hall angle of Pt, the large difference of SMR depending on the FM materials indicates that the spin Hall effect is not the sole origin of the SMR effect but possible significant contribution of FM or FM/NM interfaces[27], for example, interfacial Rashba effect[28,29] or magnetic proximity effect[30]. The latter is less likely in the W-based samples because W is far from the Stoner instability. However, further study is required to clarify the origin of the SMR.

We demonstrate in this work that the SMR experiments can be employed to extract the spin Hall angle and spin diffusion length of the NM materials in the FM/NM structures, which are essential to interpret various spin transport phenomena related to the spin-orbit coupling. Unlike other methods in which the spin pumping (or excitation) is involved, this SMR measurement allows one to obtain those parameters with a simple electrical measurement.

Lastly, the similar trend of SMR as the magnetization switching efficiency induced by the spin-orbit torque (SOT) confirms the same origin of both effects. Since SOT has been intensively investigated because of its high potentials for device applications such as current-induced magnetization switching and domain wall motion with high speed, the close examination of SMR can be very useful for understanding the SOT physics.

**Methods**

The samples of W/Co$_{32}$Fe$_{48}$B$_{20}$(CoFeB)/MgO structure were grown on thermally oxidized silicon substrates by DC/RF magnetron sputtering at Ar pressure of 3~10 mTorr. Here, the W and CoFeB thicknesses are ranged 2~7 nm and 0.8~1.4 nm, respectively. An additional Ta (1 nm) capping layer on top of MgO (1.6 nm) was deposited to prevent the contamination of the MgO layer. After deposition, thin films were annealed at 250 ℃ for 30 min in vacuum condition (less than 10$^{-5}$ Torr), which enhances the perpendicular magnetic anisotropy. Hall bar patterned devices for transport measurement (Fig. 1(a)) were fabricated using photo-lithography and Ar ion milling. The length and width of the Hall bar structure is 75 μm and 5 μm, respectively. The longitudinal ($R_{xx}$) and transverse resistance ($R_{xy}$) were measured simultaneously using a DC current of 50 μA while sweeping in-plane magnetic field or rotating the sample in the *x-y*, *y-z*, or *x-z* planes in a magnetic field which is much larger than the anisotropy field. The current-induced spin-orbit torques in the same samples were studied by performing the switching experiments. The magnetization is detected by measuring the AHE voltage after each current pulse of 10 μs width with application of an in-plane magnetic field ($H_x$) of 200 Oe, parallel to the current direction, for deterministic switching. All measurements were done at room temperature.

**Acknowledgements**

The authors acknowledge K.-J. Lee, Y. M. Kim and K.-D. Lee for critical reading of the manuscript. This research was supported by the National Research Foundation of Korea (NRF) funded by the Ministry of Science, ICT & Future Planning (NRF-2012R1A1A1041590, NRF-2014R1A2A1A11051344).

**Additional information**

While we were preparing the paper, we became aware that a large SMR in W-based structures was also observed by other group [Kim, J. *et al.* Giant spin Hall magnetoresistance in metallic bilayers. arXiv:1503.08903 (2015)].

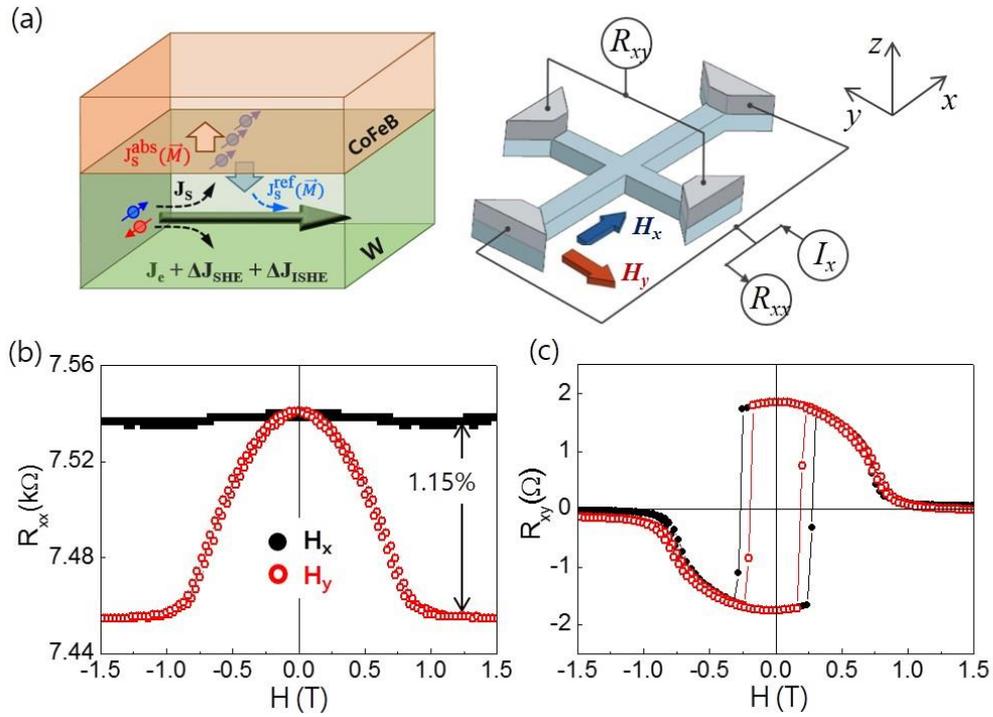

**Figure 1 | The spin Hall magnetoresistance (SMR) in W/CoFeB/MgO structure.** (a) Schematic illustration of SMR, based on the interaction of the spin current induced by the spin Hall effect with magnetization direction. The measurement scheme of longitudinal ($R_{xx}$) and transverse ($R_{xy}$) resistances is shown on the right. $R_{xx}$ (b) and $R_{xy}$ (c) of the sample W(5 nm)/CoFeB(1.2 nm)/MgO(1.6 nm) as a function of $H_x$ and $H_y$, respectively. The black solid (red open) circles represent the data for $H_x$ ($H_y$).

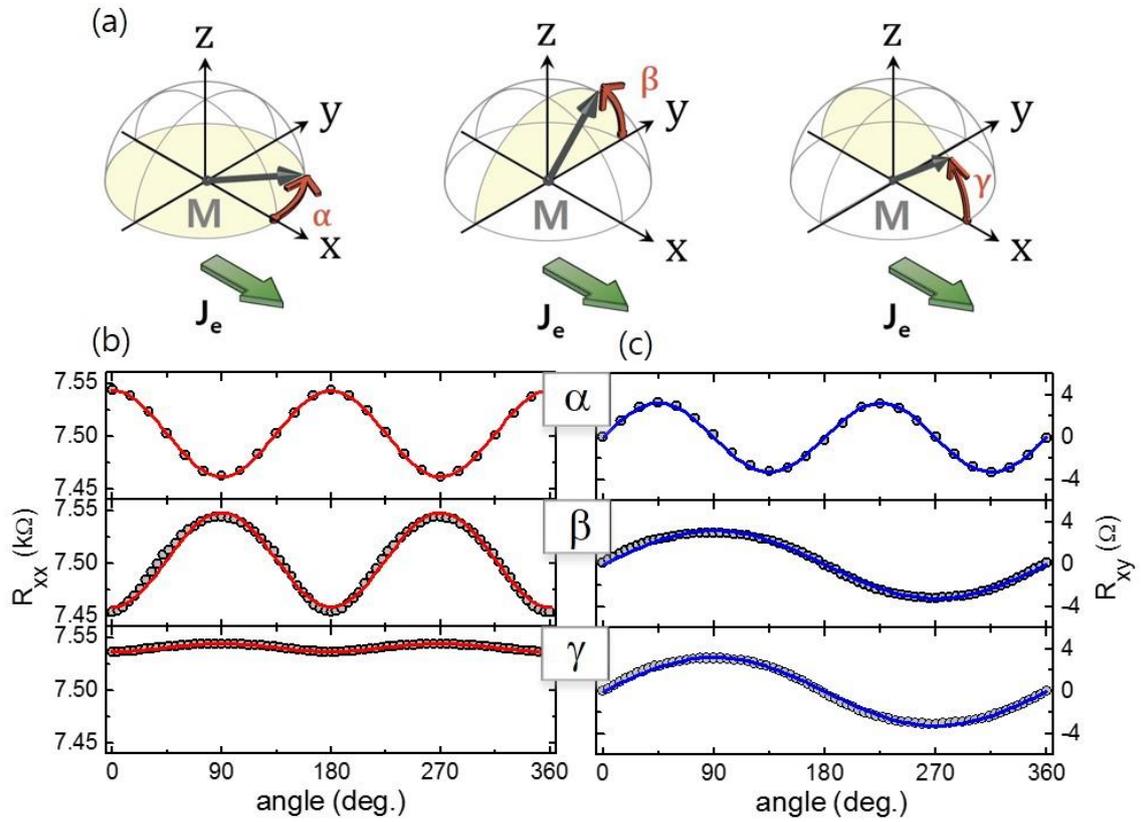

**Figure 2 | Angular dependence of the magnetoresistance.** (a) Schematic of the MR measurement using the rotating sample in a strong magnetic field, of which angles are designated as α, β, and γ, respectively. $R_{xx}$ (b) and $R_{xy}$ (c) as a function of the rotating angle α, β, and γ. The measurements were done by rotating samples in a magnetic field, 1.5 T for α-rotation and 8 T for β- and γ-rotation.

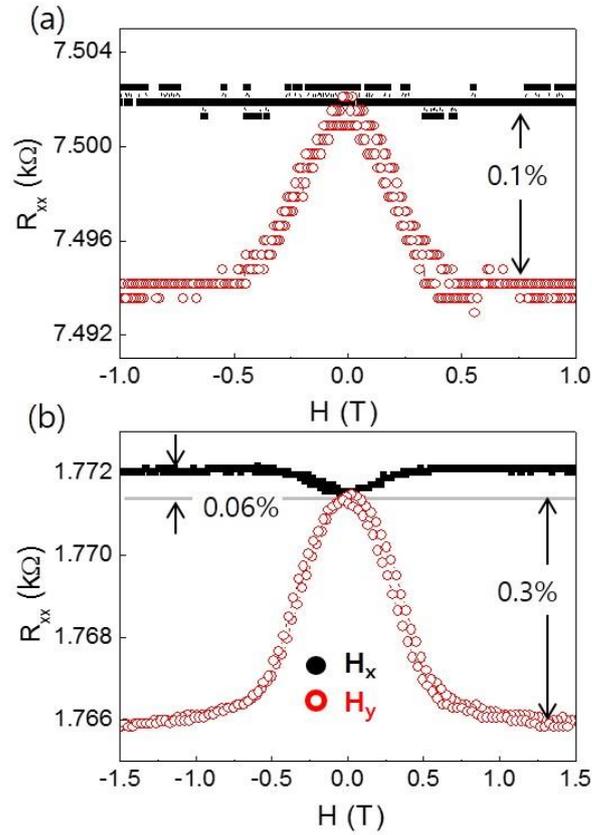

**Figure 3 | Longitudinal magnetoresistance ($R_{xx}$) for different NM underlayer.** (a) Ta(5 nm)/CoFeB(1 nm)/MgO(1.6 nm) and (b) Pt(3 nm)/Co(1 nm)/AlOx(1.5 nm) samples. The black solid (red open) circles represent the data for $H_x$ ($H_y$).

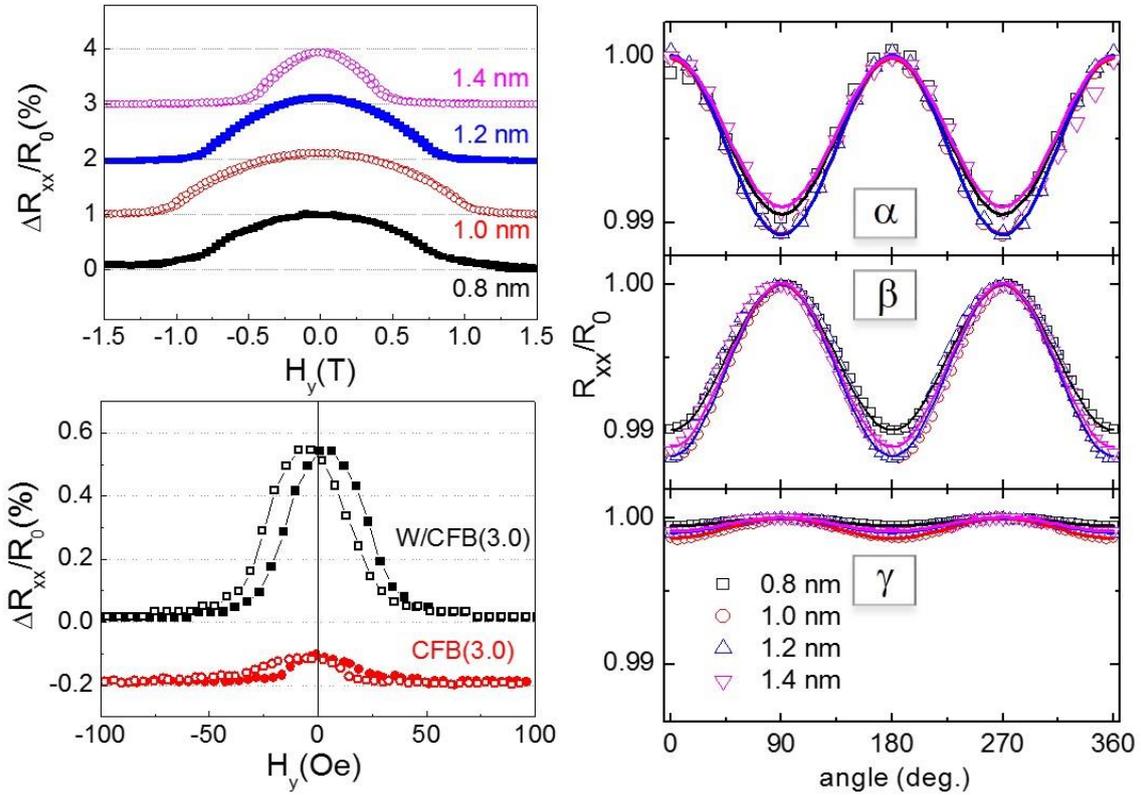

**Figure 4** | **Dependence of the magnetoresistance on the CoFeB thickness.** The MR for the sample W(5 nm)/CoFeB(*t*)/MgO(1.6 nm), where *t* varies from 0.8 to 1.4 nm using field sweep (a) and rotating (b) measurement. (c) $R_{xx}$ vs $H_y$ curves for the samples W(5 nm)/CoFeB(3 nm) and CoFeB(3 nm). Both samples show the in-plane magnetic anisotropy.

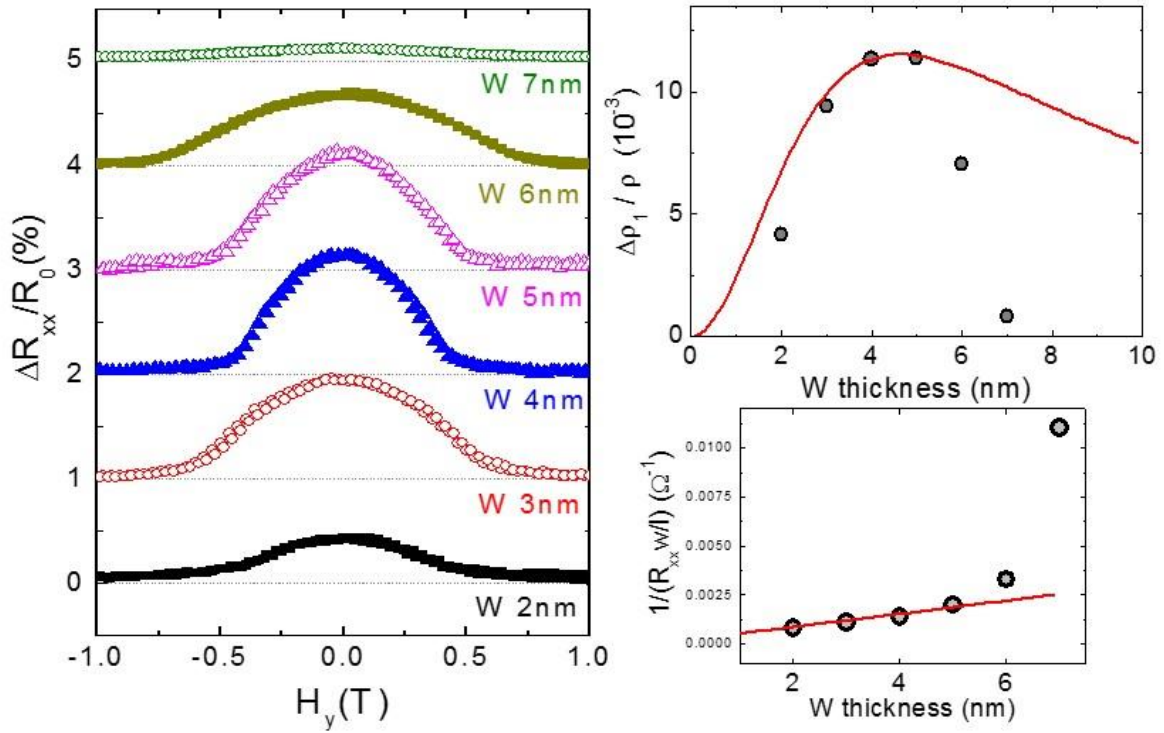

**Figure 5 | Dependence of the magnetoresistance on the W thickness.** (a) MR vs $H_y$ for the sample W(*t*)/CoFeB(1 nm)/MgO(1.6 nm), where *t* varies from 2 to 7 nm. (b) SMR as a function of the W thickness together with a theoretical fitting curve. (c) $1/R_{xx}$ vs the W thickness, from which the resistivity of W is extracted.

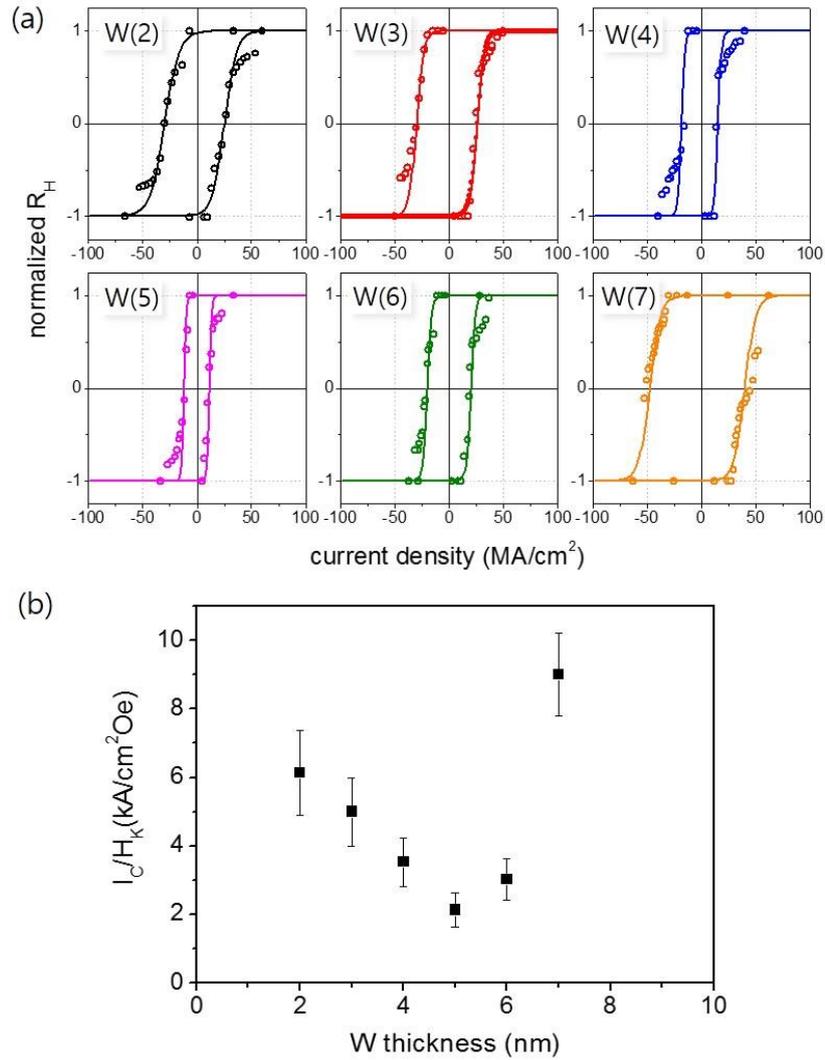

**Figure 6** | **Switching experiments utilizing spin-orbit torque induced by in-plane current.** (a) The magnetization direction detected by AHE measurement after each pulsed current of 10μs while sweeping current. The in-plane magnetic field $H_x$ of 200 Oe is continuously applied during the measurement. The lines are for a guide of an eye. (b) The ratio of critical current density ($I_c$) to magnetic anisotropy ($H_k$) as a function of the W thickness. The ratio is inversely proportional to the SOT-switching efficiency or the magnitude of SOT.